\documentclass[12pt]{article}
\usepackage{epsfig}

\makeatletter

\renewcommand{\ps@plain}{%
        \renewcommand{\@evenhead}{}%
        \renewcommand{\@oddhead}{}%
        \renewcommand{\@evenfoot}{\hfil\small{\textbf\thepage}\hfil}%
        \renewcommand{\@oddfoot}{\@evenfoot}}%

\renewcommand\section{\@startsection {section}{1}{\z@}%
                                   {-3.5ex \@plus -1ex \@minus -.2ex}%
                                   {2.3ex \@plus.2ex}%
                                   {\reset@font\bfseries}}
\renewcommand\subsection{\@startsection{subsection}{2}{\z@}%
                                     {-3.25ex\@plus -1ex \@minus -.2ex}%
                                     {1.5ex \@plus .2ex}%
                                     {\reset@font\itshape}}
\newlength{\capsize}
\renewcommand{\@makecaption}[2]{%
  \vskip\abovecaptionskip
  \sbox\@tempboxa{\small{\bfseries #1}\/: #2}%
  \ifdim \wd\@tempboxa >\capsize
   {\advance\leftskip by 0.1\textwidth \advance\rightskip by 0.1\textwidth
	{\small {\bfseries #1}: #2}\par}
  \else
    \hbox to\hsize{\hfil\box\@tempboxa\hfil}%
  \fi
  \vskip\belowcaptionskip}

\makeatother

\newcommand{\cfig}[2]{\hbox to
\columnwidth{\hfil\epsfig{file=#1,height=#2}\hfil}}

\begin{document}

\title{\large Propagating Spin Modes in Canonical Quantum Gravity}

\author{ \normalsize{\bf Roumen Borissov\thanks{Present address: Bios
Group L.P., Santa Fe, NM 87501, roumen.borissov@biosgroup.com} \ \ and
Sameer Gupta\thanks{gupta@phys.psu.edu}}\\ \\ \normalsize {\it Center
for Gravitational Physics and Geometry} \\ \normalsize {\it The
Pennsylvania State University}\\ \normalsize {\it University Park, PA
16802} } 

\date{}

\maketitle

\vspace{0.5cm}

\begin{abstract}
One of the main results in canonical quantum gravity is the
introduction of spin network states as a basis on the space of
kinematical states. To arrive at the physical state space of the
theory though we need to understand the dynamics of the quantum
gravitational states. To this aim we study a model in which we allow
for the spins, labeling the edges of spin networks, to change
according to simple rules. The gauge invariance of the theory,
restricting the possible values for the spins, induces propagating
modes of spin changes. We investigate these modes under various
assumptions about the parameters of the model.

\end{abstract}

\newpage
\section{Introduction}

There are two very important issues in the loop approach to
non-perturbative canonical quantum gravity\cite{aa86,aa92,cr91,ls92},
which we have to address in our quest for a complete theory. The first
is the problem of evolution of the gravitational quantum states and
the second one is the recovery of the classical continuous space from
the discrete structures which appear at very short scale. The problem
of evolution in canonical quantum gravity is quite non-trivial. The
underlying general covariance principle of the theory leads not to a
true Hamiltonian, but to a sum of constraints. The physical states of
the theory are then the ones which are annihilated by these
constraints. Imposing the gauge and the diffeomorphism constraints has
led to a Hilbert space whose basis is given by spin networks
space. The Hamiltonian constraint, which generates infinitesimal
translations in the time direction, has to be imposed in this space of
states in order to obtain the physical space of states.

There are different approaches which have been proposed in order to
study the action of the Hamiltonian constraint. In the canonical
approach, one relies on regularizing the Hamiltonian constraint as a
quantum operator and investigating its action on the states of the
theory. Various regularization schemes have been suggested\cite{RoSm,
Bor}, the most commonly accepted being the one proposed by
Thiemann\cite{tt96}. However, all of these suffer from various
problems (see for example \cite{ls96,gmmp97}). Another approach which
has been suggested is to look at the finite action of the Hamiltonian
constraint as opposed to the infinitesimal one\cite{rr97}. This has
led to the study of spin foams.

An alternative approach is to find a consistent set of rules for
quantum evolution which, in the classical limit, recover the action of
the Hamiltonian constraint\cite{ms97}. It is this approach that we
study in this paper. While the idea is very intriguing, it is highly
non-trivial to define an appropriate set of rules which lead to a good
classical limit. The problem of finding the appropriate rules for
evolution of the kinematical states is now related to the problem of
defining the classical limit of the theory. On the other hand the
investigation of the classical correspondence is also quite
subtle. One of the main results in canonical quantum gravity is that
geometric operators such as area and volume have discrete
spectra\cite{rs95b}. This means that space has a discrete structure at
the Planck scale. Thus, in taking the classical limit, we should also
recover the continuum space-time we see around us.

According to a proposal in \cite{ms97}, the problem of recovering the
continuum space is similar to some problems of non-equilibrium
critical phenomena in statistical physics. The long range behavior
characteristic of classical general relativity could arise at the
critical values of some parameters of the discrete theory. This is
already seen in certain discrete approaches to quantum gravity such as
dynamical triangulations. The important aspect of the proposal is
that, since there is no external agent to tune the parameters of the
universe, discrete space should be a self-organizing critical
system. The idea of self-organized criticality was introduced by Bak,
Tang and Wiesenfeld in \cite{btw87} (for reviews see
\cite{pb96,hj98}). The self-organized criticality paradigm is that
certain systems, consisting of many parts interacting via local rules,
self-organize themselves in a critical state. Usually this critical
state is characterized by the lack of specific scale of propagation of
perturbations.

In this paper, we study various models whose dynamics is suggested by
canonical quantum gravity. There are certain aspects of the action of
the Hamiltonian constraint, which are common to all of the proposed
regularizations. These aspects guide us in the introduction of rules
for graphical evolution of the spin networks. With the evolution rules
in hand, we develop a model along the lines of the sandpile models of
self-organized criticality. First we briefly introduce the spin
network states. In section 3, we discuss the various regularizations
of the action of the Hamiltonian constraint and the corresponding
graphical rules for evolution. We then introduce our model and discuss
our results.

\section{Spin networks and dual simplicial triangulations}

The basic variables in canonical quantum gravity can be taken to be an
SU(2) connection $A_{a}^{i}(x)$ and a densitized triad
$\widetilde{E}_{i}^{a}$ which takes values in the dual of $su(2)$. A
convenient basis for the space of gauge invariant states is given by
the spin network states, introduced in canonical quantum gravity by
Rovelli and Smolin\cite{rs95a} and by Baez\cite{jb95}. A spin network
state $\Gamma \equiv (\gamma, \vec{j}, \vec{I})$ is defined by the
following elements:
\begin{itemize}
\item A closed graph $\gamma$, embedded in the spatial manifold
$\Sigma$, consisting of finite number of oriented edges $e_1, e_2,
\dots, e_n$ incident at vertices $v_1, v_2, \dots, v_m$. A vertex is
called $p$-valent or of valence $p$ if there are $p$ edges incident at
that vertex.
\item Each edge $e_i$ of the graph $\gamma$ is labeled by an irreducible representation $j_i$ of SU(2). Further on, $c_i = 2j_i$ is referred to as the color of the edge.
\item At each vertex $v_j$ there is an intertwining operator from the tensor product of the representations carried by the incoming edges to the tensor product of representations labeling the outgoing edges.
\end{itemize}

When considering only the combinatorial aspects of a spin network
state it is very convenient to think of the spin network on which the
state is based as the dual 1-skeleton of a simplicial triangulation of
the spatial manifold $\Sigma$. In this picture, a spin network is
equivalent to a colored triangulation of the spatial manifold. In the
rest of this paper, we shall consider a spin network or its dual
triangulation as equivalent objects.

\section{Action of the Hamiltonian constraint}

\begin{figure}
\cfig{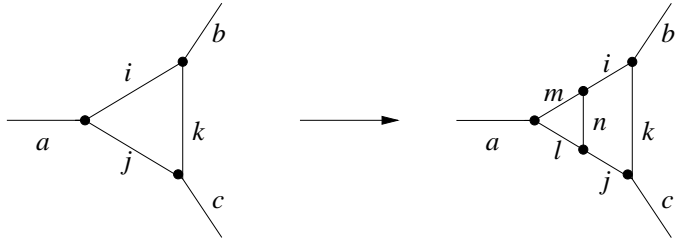}{1.25in}
\caption{Thiemann's regularization}
\label{ham1}
\end{figure}

We shall now schematically describe the graphical action of the
Hamiltonian constraint of canonical quantum gravity. Essentially the
action of the Hamiltonian constraint can be thought of as adding a
loop to the original spin network with some extra requirements. First,
the added loop should coincide with two of the edges of the original
spin network joined at one vertex in the vicinity of that
vertex. Second, the added loop can coincide with any other (portions
of) edges. Whenever the added loop runs along already existing edge,
the recoupling theory of angular momentum determines the way the
original spins change.

\begin{figure}
\cfig{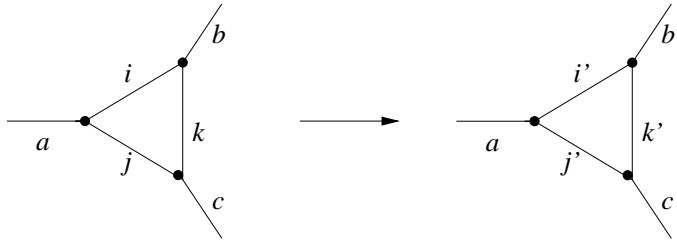}{1.25in}
\caption{Smolin's regularization}
\label{ham2}
\end{figure}

According to the regularization described in \cite{tt96}, the
Hamiltonian acts only in a vicinity of the vertices of the spin
network. For simplicity let us consider a trivalent vertex. At such a
vertex the action of the Hamiltonian constraint can be thought of as a
sum of three terms, each one corresponding to the old spin network
with an added new edge of color one connecting two of the original
edges (see Figure \ref{ham1}). The colors of the original edges in
their portion between the newly created vertices and the original
vertex are either increased or decreased by one. Consecutive action of
the Hamiltonian constraint keeps adding new edges only to the original
ones, thus producing a web-like structure at each vertex. The problem
with such a definition of the action of the Hamiltonian constraint is
that it is ultra-local in the sense that the action of the constraint
at one vertex does not propagate any information to the neighboring
vertices of the spin network. Hence, there are no long range
correlations.  A modification which was aimed at fixing this problem
was introduced in \cite{ls96}. This new version of the constraint
action allows not only the addition of new edges but also the addition
of whole loops coinciding completely with the already present
edges. In such cases the original graph does not change at all but the
edges of coincidence have their colors changed. This is shown in
Figure \ref{ham2}.

In our work, we model the action of the Hamiltonian constraint by
introducing a generalization of this second type of regularization. We
assume that loops can be added not only to a single closed cycle of
edges, but also can run along any collection of the original
edges. Also, a very important assumption is that the added loop can
run many times along one and the same edge. Since the effect of adding
a loop segment to an existing edge simply changes the color of that
edge, we can move away from the loop picture and consider simply gauge
invariant changes of the colors of the edges of the original spin
network.

\section{The Model}

\begin{figure}[h]
\cfig{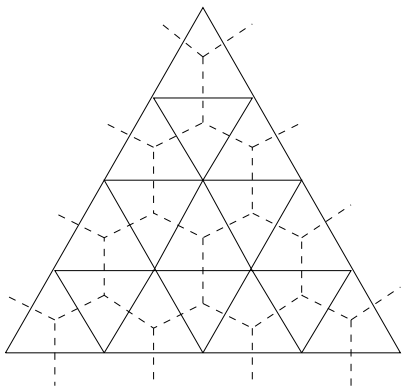}{2in}
\caption{The triangulated triangle and its dual graph}
\label{space}
\end{figure}

For simplicity, we study only planar trivalent spin network states in
a 2-dimensional space with boundary (topology of a disk). It is
convenient to take this space to be a triangle. The spin networks
associated with that triangle are based on the graph, dual to the
triangulation of the triangle. The triangulation is performed by
subdividing the sides of the triangle into $N$ parts. The triangulated
triangle is shown in Figure \ref{space}. The dual graph on which the
spin network states are based is obtained by assigning a vertex to
each 2-cell in the triangulation and an edge crossing each 1-cell and
connecting the vertices in the neighboring 2-cells (shown using dashed
lines in Figure \ref{space}). The colors carried by the edges of the
spin network can be thought of as being lengths of the corresponding
crossed 1-cells of the triangulation\cite{cr94}. More precisely, the
length of a side of the triangulation which is intersected by a spin
network edge of color $c$ is
\begin{equation}
l = \frac{c}{2}l_\mathrm{P}
\end{equation}
where $l_\mathrm{P}$ is the Planck length\footnote{Recall that in 2D,
$l_\mathrm{P} = \hbar G/c^3$.}. Thus, we shall refer to the
triangulation and lengths or its dual spin network and the
corresponding colors interchangeably.

The requirement for gauge invariance at the vertices of the spin
network state is essentially a triangle inequality that the lengths of
the sides of a triangle should satisfy together with the requirement
that the sum of these lengths be an even number. We call the triangle
inequality together with the requirement for an even sum of the site's
lengths a gauge invariance condition. If $a$, $b$ and $c$ are the
colors of the three edges at a node, the gauge invariance requires
that:
\begin{eqnarray}\label{gge}
        a + b &\ge& c\nonumber\\
        b + c &\ge& a\\
        c + a &\ge& b\nonumber\\
        a + b + c &=& \mathrm{even} \label{even}
\end{eqnarray}   

Further, we will assume that the colors of the edges of the spin
network crossing the boundary are unconstrained by the evolution
process. As an initial state for our system, we construct a spin
network based on the dual graph and assign random colors to the edges
of the spin network in a gauge invariant fashion. The evolution of
spin networks in our model corresponds to changing a set of colors
such that the gauge invariance condition remains satisfied at each
node of the spin network. We use the following procedure to implement
this evolution. Choose, at random, one of the edges of the spin
network and change its color by an amount $\Delta c$. This will, in
general, affect the gauge invariance of the two nodes on which the
edge is incident. To restore the gauge invariance at the nodes, which
are no longer gauge invariant, we change the color of one of the
remaining incident edges. We continue this process until gauge
invariance is restored at every node of the graph.

Depending upon the value of $\Delta c$, there are various
possibilities for the exact implementation of the described
procedure. For an odd $\Delta c$, say $\Delta c = \pm 1$, the gauge
invariance condition is always violated at the two nodes on which the
initial edge is incident, because of (\ref{even}). The restoration of
the gauge invariance of the first pair of nodes will necessary induce
violation of the gauge condition at two new nodes. There are two cases
for this pattern to end -- either at a certain step a non-invariant
node contains an edge which crosses the boundary of the triangle or
the spin changes meet somewhere on the spin network and neutralize
each other. These two cases can be viewed from a loop perspective as
corresponding to -- (i) adding an open line of color 1, whose both
ends lie on the boundary or (ii) adding a closed loop of color 1.

\begin{figure}[b]
\cfig{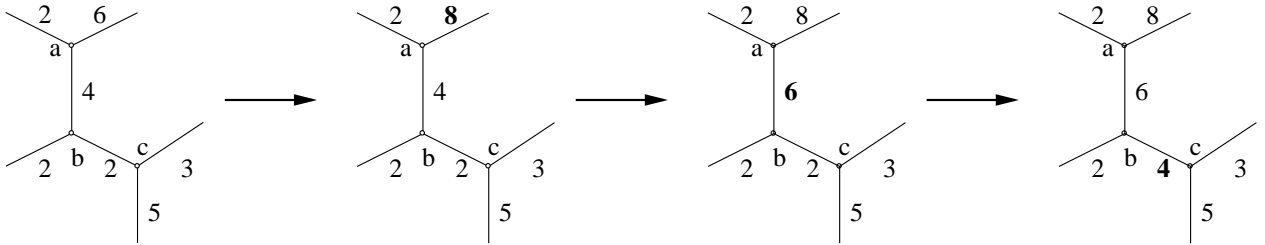}{1.25in}
\caption{An example of an avalanche of spin changes. The initial
perturbation changes the color of the color 6 edge incident at node
`a' to 8. This makes the node gauge non-invariant and forces the color
4 edge to change to color 6. This in turn affects node `b' and so on.}
\label{aval}
\end{figure} 

The case of even $\Delta c$, say $\Delta c = \pm 2$, is more
involved. It is possible that the change of the color of a particular
edge by $\pm 2$ leaves both the original nodes gauge
invariant. Actually this is true in most of the cases -- requirement
(\ref{even}) is not violated and it is up to the conditions
(\ref{gge}) to determine the gauge invariance. We will mostly be
interested in this latter case.

Let us now discuss the case $\Delta c = \pm2$ in more detail. Let us
choose an edge of the graph at random and change the length of this
edge by $\pm2$. The only case in which this can violate gauge
invariance is if the triangle was ``flat'' initially, that is, if one
of the three inequalities in equation (\ref{gge}) is saturated. We
call such triangles (vertices) critical. If, for such a critical
triangle, we increase the longest side or decrease one of the two
shorter sides, we will violate gauge invariance. To compensate for
this, we have to adjust the length of one of the other two sides. The
adjusted side also belongs to one of triangle's neighbors. If this
change violates the gauge invariance condition in the neighboring
triangle, we adjust the colors there as well. This process will
continue until all the triangles in the triangulation satisfy gauge
invariance again. Thus, one change of $\pm2$ can lead to a series of
spin changes (see Figure \ref{aval} for an example). These spin
changes are analogous to the topplings in the sandpile models. Drawing
further analogy with the sandpile models, we refer to the series of
spin changes required to restore gauge invariance after the initial
change as an avalanche.

Since we allow arbitrary spins to start with, the number of critical
triangles will be negligible in comparison to the total number of
triangles. Thus if we change the spins in an arbitrary way almost no
avalanches will occur. In the sandpile model the analogous situation
is when we add and remove grains of sand with equal
probabilities. Hence, we introduce a probability $p$ for ``flattening"
of a triangle. If the arbitrary chosen side has the biggest length in
the triangle, we increase the length by 2 with a probability $p$ and
decrease it with a probability $1-p$. Similarly if the arbitrary chose
side is the smallest one or the one in the middle, we decrease the
length by 2 with probability $p$ and increase it with probability
$1-p$. Thus it is clear that for $p$ close to 1, we will be pushing
the triangles in the triangulation towards criticality. The rest of
the avalanche (if any) proceeds a deterministic way as described
above. As a further modification of the above update rule, we
introduce the probabilistic update into the entire avalanche. Then,
there is a probability $p$ that a gauge non-invariant node is restored
to gauge invariance and probability $1-p$ for it to go further away
from gauge invariance.

We performed simulations for system sizes ranging from 400 to 10,000
triangles with a typical run covering a million iterations of the
system.

\section{Results and Interpretation}

\begin{figure}
\cfig{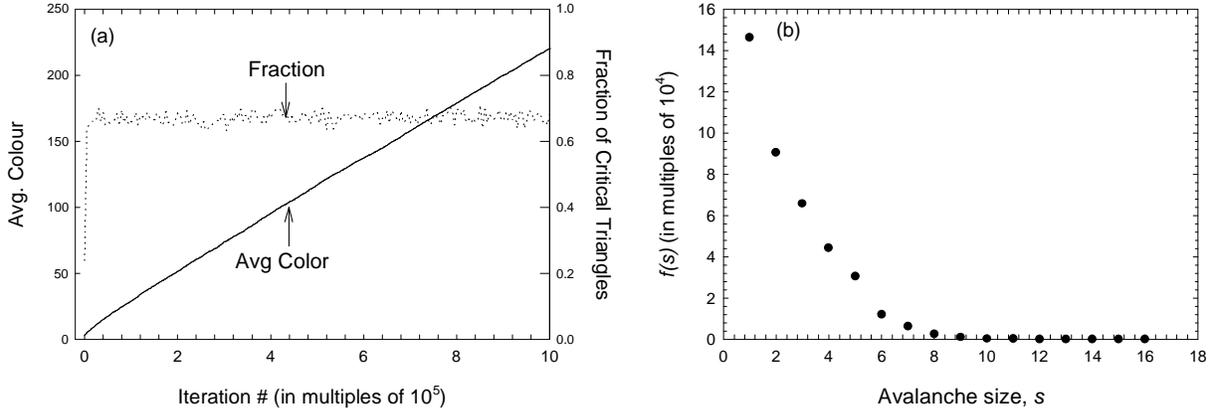}{2.15in}
\caption{Results for a typical run with 400 triangles. Graph (a) shows
how the average color per edge and the fraction of critical triangles
in the system evolve as the simulation proceeds. Graph (b) shows the
distribution of avalanches $f(s)$ as a function of size.}
\label{t20-2}
\end{figure}

To start with, we simulated the system with the deterministic
evolution rule for system size of 400 triangles. The results of these
simulations are shown in Figure \ref{t20-2}. As can be seen from graph
(a) in the figure, the fraction of critical triangles in the system
rises very rapidly from an initial value of 0.24 to reach an
equilibrium value of around $0.67\pm0.01$. The average color per edge
in the system increases linearly as the simulation proceeds.

The distribution of avalanches $f(s)$ as function of avalanche size,
$s$ is shown in graph (b) of Figure \ref{t20-2}. This distribution can
be described by an exponential decay
\[
f(s) \sim \exp(-s/\sigma)
\]
to a good accuracy.

\begin{figure}[t]
\cfig{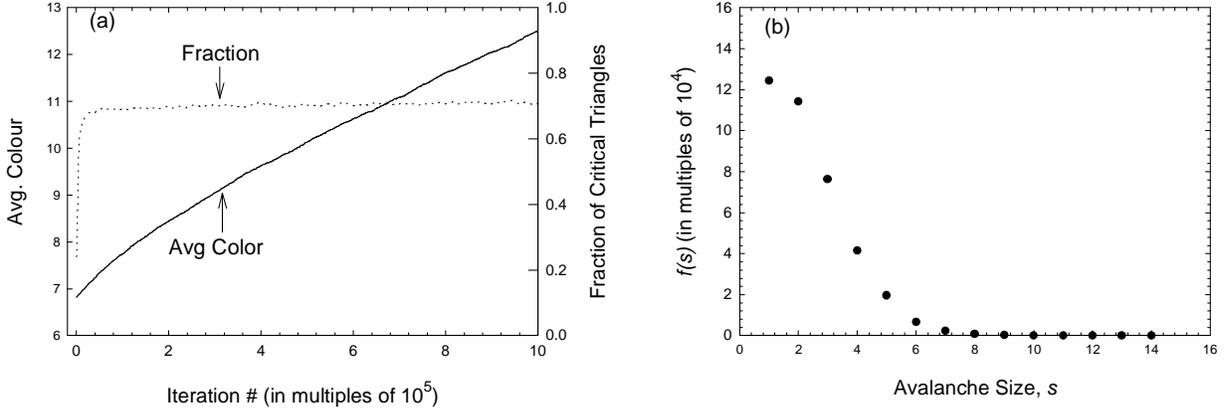}{2.15in}
\caption{Results for a typical run with 10,000 triangles. Graph (a)
shows how the average color per edge and the fraction of critical
triangles in the system evolve as the simulation proceeds. Graph (b)
shows the distribution of avalanches $f(s)$ as a function of size.}
\label{t100-2}
\end{figure}

The results for a larger system size (10,000 triangles) are shown in
Figure \ref{t100-2}. The fraction of critical triangles reaches its
equilibrium value more slowly as compared to the smaller system. The
equilibrium value of $0.70\pm0.005$ is similar to that for 400
triangles. As expected, the average color per edge increases at a
slower rate as well. The avalanche distribution can again be described
well by a decaying exponential. Thus, our results scale with the
system size.

The simulations for the case where we introduce the probability $p$
into the first move were performed next. Depending upon the value of
$p$, a certain fraction of the triangles in the system become
critical. Once reached, the fraction of critical triangles remained
almost constant. The avalanche distribution was again described by an
exponential decay with the decay constant decreasing with decreasing
probability. The data remained qualitatively similar to Figures
\ref{t20-2} and \ref{t100-2}.

\begin{figure}[b]
\cfig{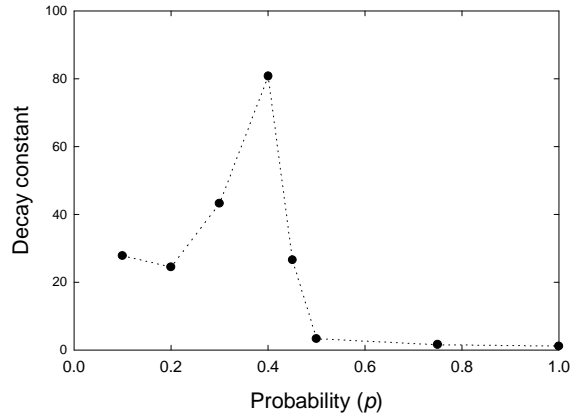}{2.15in}
\caption{Behaviour of the decay constant as a function of probability}  
\label{prob}
\end{figure}

Finally, we simulated the case where the probabilistic update was also included in the avalanche phase of the dynamics. The frequency distribution of the avalanches was still described by an exponential decay. The decay constant however did not decrease as we decreased the probability from 1. It initially increases till $p = 0.4$ and then decreases again (see Figure \ref{prob}). 

Figure \ref{prob} seems to indicate that our system undergoes a phase
transition around $p = 0.4$. However, if we look at the avalanche
distribution more carefully, we find that while we get some extremely
large avalanches, the number of sites involved in these avalanches is
very small. For example, the largest avalanche that we got for the 400
triangles case consisted of 819 topplings which affected only 12
edges. This implies that while we seem to have a phase transition, it
does not lead to long range correlations.

\section{Conclusions}

We developed a model of evolving spin networks in which we keep the
underlying graph fixed and consider changes of the colors of the
edges. In the case when the colors are changed by $\Delta c=\pm1$,
these changes involve either sets of edges which form closed cycles or
run between the boundaries of the space. When the colors are changed
by $\Delta c = \pm2$, the response of the system is more involved. Our
aim was to prescribe a set of rules for the spin changes and then
check if such spin changes would propagate in a scale free behavior,
characteristic for a SOC system. The answer to this question to a very
large extent depends on the specific features of the model we
proposed. It turned out that with the rules we proposed, the system
does not exhibit SOC behavior. Instead we found that for particular
values of the parameters in the system we got a phase transition. At
this point, it cannot be decided using a different set of rules will
lead to the kind of self-organized behaviour that we were
seeking. There are various possibilities for modification of the
evolution rules which we are exploring currently. These include:

\begin{itemize}

\item Mixed color updates --- for each vertex, at which the gauge
condition is violated by $\Delta c=\pm2$ introduce with certain
probabilities gauge invariance restoration either by update of the
color of one of the edges by $\Delta c=\pm2$ or of two of the edges by
$\Delta c=\pm1$.

\item Conserved total color --- after adding a color of 2 to a
particular edge, we continue by redistributing the added color among
the adjacent edges, thus preserving the total color of the spin
network.

\item One directional propagation --- starting from an edge, one of
the nodes of which remains gauge invariant after the spin change. This
will avoid the problem of order of overlapping present in the case
when the spin changes propagate in two directions.

\item Spin changes starting at the boundaries and propagating into the
system.
\end{itemize}

Further work is needed to show if an appropriate set of evolution
rules can result in the definition of the spin network states as a
self-organizing critical system.

\section*{Acknowledgements}

We would like to thank Lee Smolin for the helpful discussions, which
led us to this paper and for his important comments. We are also
grateful to Fotini Markopoulou and Maya Paczuski for the useful
comments and suggestions. This work was supported in part by NSF grant
PHY-9514240 to the Pennsylvania State University.

\end{document}